# Momentum-Resolved Electronic Structure and Orbital Hybridization in the Layered Antiferromagnet CrPS$_4$

Lasse Sternemann[1,*,+], David Maximilian Janas[1,#,+], Eshan Banerjee[2], Richard Leven[1], Jonah Elias Nitschke[1], Marco Marino[1], Leon Becker[3], Ahmet Can Ademoğlu[1], Frithjof Anders[1], Stefan Tappertzhofen[3], and Mirko Cinchetti[1].

[1]*TU Dortmund University, Department of Physics, 44227 Dortmund, Germany.*

[2]*Department of Materials, Imperial College London, London, SW7 2AZ, United Kingdom.*

[3]*TU Dortmund University, Department of Electrical Engineering and Information Technology, 44227 Dortmund, Germany.*

*Corresponding authors:* * lasse.sternemann@tu-dortmund.de; # david.janas@tu-dortmund.de.

*Note:* + *these authors contributed equally*

**Abstract**

Chromium thiophosphate (CrPS$_4$) is a layered two-dimensional antiferromagnetic semiconductor exhibiting intriguing spintronic and magneto-optical properties, yet its electronic band structure has remained experimentally uncharacterized. Here, we employ momentum-resolved photoemission spectroscopy above and below the Néel temperature, complemented by density functional theory with Hubbard U corrections (DFT+U), to reveal a valence band dominated by Cr 3$d$ and S 3$p$ states with a ligand-to-metal charge-transfer band gap. We identify weakly hybridized t$_{2g}$ orbitals responsible for magnetic ordering and strongly hybridized e$_g$ orbitals that relax dipole selection rules, enabling optically active orbital transitions. These findings establish a foundational understanding of CrPS$_4$'s electronic structure, providing a benchmark for theoretical models and informing future investigations into its orbital physics and potential device applications.

**Introduction**

The first exfoliation of graphene in 2004[1] triggered intensive research on two-dimensional (2D) van-der-Waals (vdW) materials. Today, vdW materials serve as a tunable platform for exploring new physical phenomena and device concepts[2–4]. Among these systems, magnetic 2D materials are particularly compelling as they enable spin-dependent functionalities down to atomic thicknesses[5–9]. Following the discovery of ferromagnetism in a single layer of $CrI_3$[10], research has expanded toward antiferromagnetic vdW materials, which offer distinct technological advantages: the absence of stray fields enables dense integration, while high-frequency magnon modes reaching the terahertz regime and orbital excitations on the femtosecond scale open routes to ultrafast information processing[11–13]. Recent demonstrations include giant magnetoresistance (GMR) in twisted bilayers of CrSBr[14] or photoluminescence enhancement in $MoS_2$/$FePS_3$ heterostructures[15], highlighting the versatility and potential of this material class.

Chromium thiophosphate ($CrPS_4$) has emerged as a particularly promising vdW material[16,17]. It is a layered antiferromagnetic semiconductor with a Néel temperature of 38 K, characterized by ferromagnetic intralayer and antiferromagnetic interlayer coupling[18–20]. Optical studies identified pronounced sub-gap *d-d* transitions near 1.6 eV and 1.8 eV[21–23], which not only define the optical fingerprint of $CrPS_4$ but can also serve as efficient channels for the generation of coherent phonons and magnons[11]. Such strong coupling between orbital excitations and collective lattice or spin dynamics renders $CrPS_4$ as an appealing platform for magneto-optical and spintronic applications across the visible spectrum. This potential is further underscored by evidence for GMR[24] and magnon-magnon coupling[25].

Despite extensive research into its magnetic, optical, and transport properties[26–29], no experimental analysis of the electronic band structure of $CrPS_4$ has been reported. Although numerous density functional theory (DFT+U) calculations have been reported[23,24,28,30,31], these theoretical results lack an experimental benchmark, limiting the ability to directly validate theoretical simulations.

In this work, we close this gap by presenting the first experimental electronic band structure analysis of $CrPS_4$, measured above and below its Néel temperature using angle-resolved photoemission spectroscopy (ARPES), complemented by DFT+U calculations. To overcome the charging effects that typically hinder photoemission studies of insulating vdW crystals, we exfoliated $CrPS_4$ multilayers onto a conductive gold thin-film substrate, ensuring reliable spectral acquisition at both room and cryogenic temperatures. The measured spectra show excellent agreement with DFT+U calculations for the antiferromagnetic phase and reveal no significant differences across the magnetic transition within the experimental resolution.

Building on theoretical insights, we analyze the Cr 3*d* manifold, focusing on two distinct hybridization regimes between Cr 3*d* orbitals and Sulfur (S) 3*p* orbitals. Weakly hybridized $t_{2g}$ orbitals form bands broadened to 3 eV for the respective spin channels. These bands retain a strong local spin-polarized character consistent with Hund's rule filling, which relates to the magnetic order. By contrast, strongly hybridized $e_g$ orbitals form bonding-antibonding bands, which are separated by approximately 4 eV. This orbital-mixing relaxes dipole selection rules, enabling otherwise suppressed optical *d-d* transitions.

Together, these findings provide a comprehensive microscopic picture of the electronic structure of $CrPS_4$, establishing a benchmark for theoretical models and opening new perspectives for orbital engineering, magneto-optical control, and device design based on layered antiferromagnetic semiconductors.

## Results and Discussion

The crystal structure of $CrPS_4$ consists of $Cr^{3+}$ ions coordinated in distorted S octahedra that are interconnected by pairs of phosphorous (P) atoms. The S octahedra form vdW layers in the *ab*-plane, stacked along the *c*-direction, as illustrated in Figure 1A. The primitive orthorombic unit cell, superimposed onto the crystal structure in Figure 1A, displays a C2/m symmetry. Below 38 K, the intralayer ferromagnetic order consists of out-of-plane magnetic moments tilted by about 10° from the surface normal and is alternated in the adjacent layers, producing the A-type antiferromagnetic order sketched in Figure 1A[20].

In the following, we verify the structural integrity of the measured flake after continuous exposure to XUV radiation during the photoemission study, ensuring that the ARPES measurements reflect intrinsic properties. The latter will be discussed via a detailed comparison of experimental spectra at room temperature (300 K) and 10 K to DFT+U calculations, with the DFT+U results being used for further investigation of the octahedral crystal field (CF) effect on the Cr 3*d*-manifold.

*Structural Characterization:*

Photoemission measurements were only applicable for thin flakes on conductive substrates, as bulk crystals displayed significant charging under photoemission, even at room temperature, preventing the acquisition of photoemission spectra[32–34]. Therefore, the $CrPS_4$ flake was exfoliated onto a gold template, as described in the methods section and imaged in Figure 1B, C by optical microscopy and photoemission electron microscopy (PEEM). Subsequently, an investigation of the flake's thickness and surface morphology was carried out by means of atomic force microscopy (AFM). The line profile (LP) along the path highlighted in the optical image was extracted from an AFM scan presented in Supplementary Section S1. As shown in Figure 1D, it captures the step-wise transition from the gold thin-film to the flake surface, yielding a thickness of 385 nm. An additional AFM scan (Supplementary Section S1) reveals a low surface roughness (arithmetical mean height of 0.22 nm), confirming the high quality of the exfoliated crystal. Raman spectroscopy was employed to assess possible beam-induced damage during the photoemission study, probing the ARPES spot marked by a white circle in the PEEM image. Spectra taken at intentionally damaged spots display an additional peak at 466 cm$^{-1}$ (Supplementary Figure S2.1), which serves as a reference for irradiation damage. The spectrum taken at the ARPES probe spot, presented in Figure 1E and investigated in detail in Supplementary Section S2, is devoid of this peak and reproduces the characteristic peaks reported in earlier studies[21,35,36], demonstrating both high crystallinity and the absence of irradiation damage inflicted during the photoemission study.

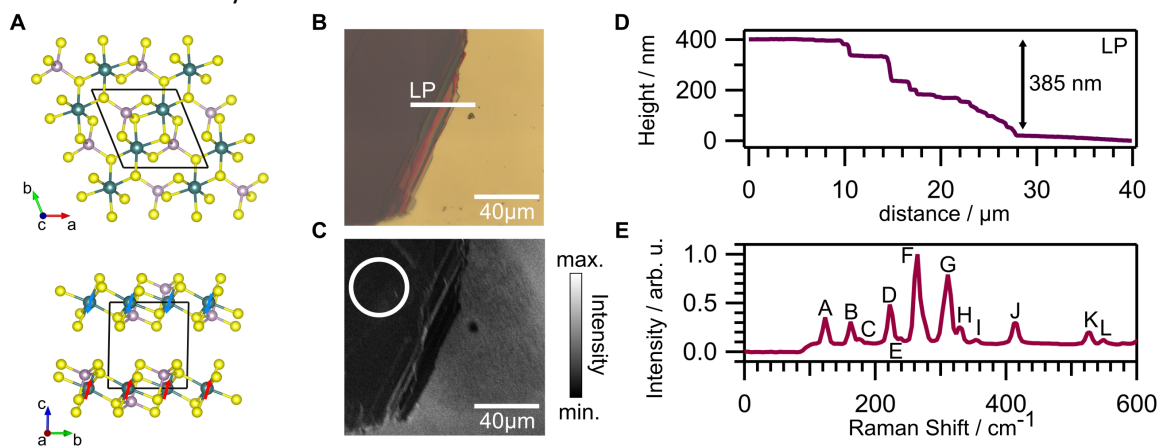

**Figure 1. Structural Properties of the CrPS$_4$ flake. A** CrPS$_4$ crystal structure as seen along the *c*- and *a*-axis, respectively. Black lines indicate the primitive unit cell, with the projection along the *c*-axis being indicative of the 1$^{st}$ Brillouin zone detectable in ARPES. **B** and **C** display the investigated flake imaged by optical microscopy and PEEM, respectively, with the circle in **C** indicating the ARPES measurement position. **D** The line profile extracted from an AFM scan along the direction highlighted in the flake's optical image indicates a sample thickness of 385 nm. **E** Raman spectrum taken after the ARPES investigation at the probed position.

*Momentum-Resolved Photoemission Spectroscopy and DFT+U:*

The experimental momentum-integrated photoemission intensity or electron distribution curve (EDC) referenced to the valence band maximum (VBM) $E_{VBM}$ is presented in Figure 2A, alongside the calculated density of states (DOS), which shows good agreement with the measured spectrum. Using the EDC, the work function was determined to $(4.9 \pm 0.2)$ eV via the secondary electron cutoff as described in Supplementary Section S3. The spin-resolved and atomic orbital-projected DOS (pDOS) in Figure 2B reveals the orbital composition underlying the electronic structure within one layer. Consistent with previous theoretical studies[29,37–39], the pDOS demonstrates a dominant contribution from S 3*p* states extending throughout the entire valence band region down to -6.5 eV and showing a slight asymmetry of spin channels reversed with respect to the Cr 3*d* states.. The major contribution of Cr 3*d* states is located between 0.0 eV and -3.5 eV, fully overlapping with the S 3*p* states. Due to the strong spin-polarization of the Cr *d* states as well as the extended overlap between Cr *d* and S *p* states, this region may pose special interest for possible hybridization and spin phenomena. The prevalence of Cr and S contributions in the electronic structure is further evident in the material's band gap, which, in agreement with prior studies[23,38], displays a ligand-to-metal charge-transfer character. In contrast to Cr and S, P *p* orbitals contributions are confined to energies below -3.5 eV and exert minimal influence on the near-VBM structure. Notably, the *s* orbitals of each element deliver no significant contribution to the probed energy region and are therefore omitted in Figure 2.

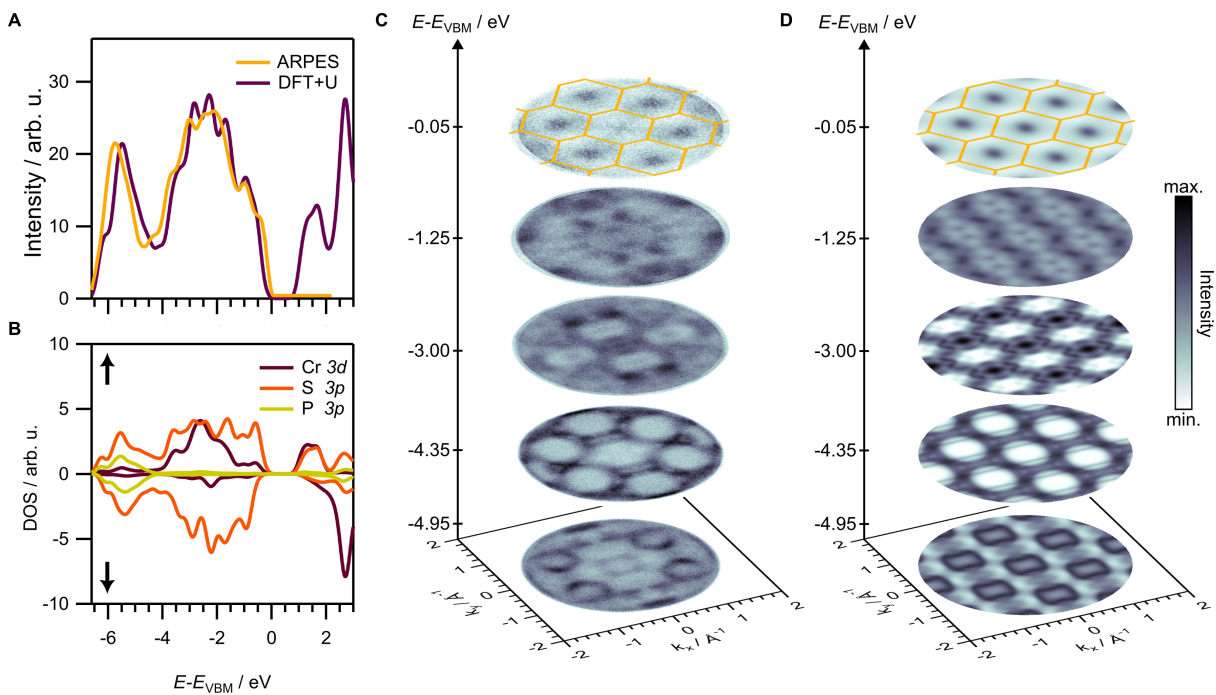

**Figure 2. DOS and Momentum Maps of CrPS$_4$. A** The experimentally obtained EDC alongside the DOS extracted from DFT+U calculations. **B** The atomic orbital projected and spin-resolved DOS displays a mixed contribution of Cr *d* and S *p* orbitals near the VBM, while the P *p* orbitals only give a minor contribution starting below -3.5 eV. **C** and **D** highlight the resemblance of the experimentally obtained momentum maps at selected energies next to their theoretical counterparts calculated using DFT+U, with both momentum maps at -0.05 eV being superimposed by the BZ boundaries. The data was acquired using linearly-polarized light with a photon energy of $h\nu = 21.21$ eV.

To enable direct comparison between momentum-dependent experimental and theoretical results, we extract isoenergetic momentum maps, i.e. intensity distributions $I = I(k_x, k_y, E - E_{VBM})$, from both ARPES measurements and DFT+U calculations. To minimize the impact of matrix element-induced intensity variations, dependent on experimental conditions[40], and enhance comparability with theory, we apply C2/m symmetry operations to symmetrize the experimental momentum maps according to the primitive unit cell symmetry. The raw momentum maps, as well as a detailed description of the symmetrization process, are presented in Supplementary Section S5. The VBM momentum map extracted at -0.05 eV, depicted in Figure 2C, displays the characteristic centered-rectangular surface Brillouin zone (BZ) boundaries (cf. Supplementary Section S4), with dominant photoemission intensity concentrated at the $\overline{\Gamma}$-points. The accurate reproduction of this intensity distribution by DFT+U calculations, shown in Figure 2D, validates the theoretical framework. Excluding matrix element effects not compensated by the symmetrization procedure, the calculated momentum maps resemble the experimental momentum distribution throughout the probed energy region and thereby all significant orbital contributions.

Figure 3A presents the momentum map at -4.10 eV with overlaid high symmetry points ($\overline{\Gamma}$, $\overline{X}$, $\overline{Y}$, $\overline{S}$). Band dispersions extracted along two principal crystallographic directions, $\overline{Y}$-$\overline{\Gamma}$-$\overline{Y}$ (cut 1) and $\overline{S}$-$\overline{\Gamma}$-$\overline{S}$ (cut 2), are put to comparison against DFT+U results in Figures 3B and C, respectively. The negative $k_{//}$-values represent experimental data, while positive $k_{//}$-values display the corresponding theoretical bands, with high symmetry points marked by dashed orange lines.

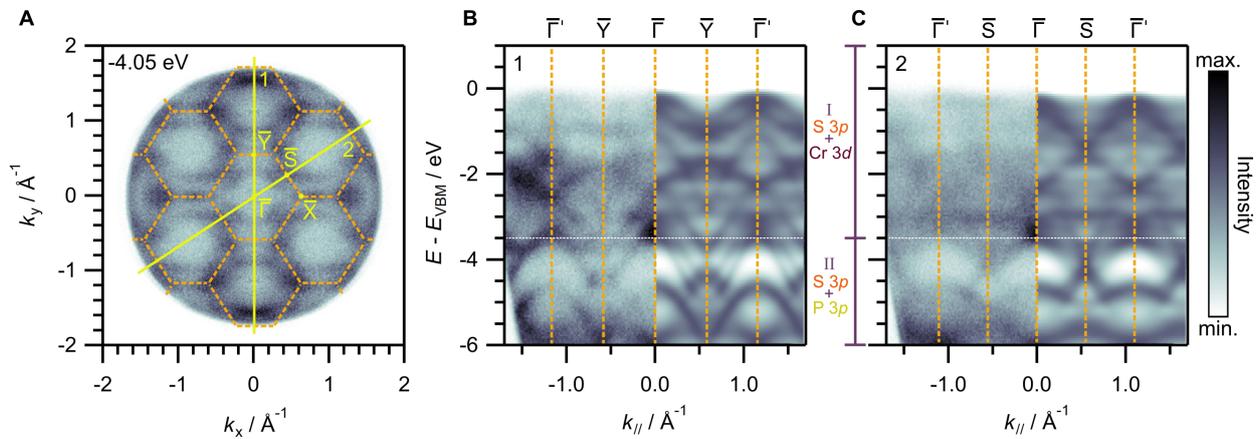

**Figure 3. Experimental and Theoretical Band Structure. A** The momentum map recorded at -4.05 eV with the BZ boundaries (orange lines) highlighted alongside the high symmetry points $\overline{\Gamma}$, $\overline{X}$, $\overline{Y}$ and $\overline{S}$ (yellow dots). Cuts extracted along directions 1 and 2 yield the experimental band structure presented alongside DFT+U calculations for the $\overline{Y}$-$\overline{\Gamma}$-$\overline{Y}$- **B** and $\overline{S}$-$\overline{\Gamma}$-$\overline{S}$-direction **C**. The experimental data are plotted for negative $k_{//}$-values, while the theory results are put to comparison at positive values. The data was acquired using linearly-polarized light with a photon energy of $h\nu$ = 21.21 eV.

In analogy to the DOS, the band structure can be separated into two regions, dominated by a mixture of Cr $d$ and S $p$ states (I: 0.0 eV to -3.5 eV) and P and S $p$ states (II: -3.5 eV to -6.0 eV), respectively. Region II is dominated by highly dispersive bands, which are replicated particularly by the theoretical simulations, enabling direct band-by-band identification. Region I displays a complex behavior governed by the overlap of multiple either highly dispersive or localized bands. While theoretical results capture most experimental features, the high density of bands hinders the direct identification of single bands. Within 1.0 eV below the VBM, the density of bands is reduced, revealing an upward dispersive behavior observable in experiment and theory along both high-symmetry directions.

As previously described, CrPS$_4$ transitions from a paramagnetic to an A-type antiferromagnetic order below 38 K. The investigation of this magnetically ordered phase, carried out at 10 K, presents significant experimental challenges, as charging effects that already require suppression via thin-flake preparation on conductive substrates at room temperature are further enhanced at cryogenic temperatures. Under photoemission at 10 K, the increased electrical resistivity induces positive surface charging, resulting in a photoelectron kinetic energy shift, which accumulates to 2 eV over 2 hours of measurement time. We carefully analyzed and calibrated the energy shift, as described in Supplementary Section S6, to correct for these charging effects. Additionally, spectral broadening from contamination at low temperatures and charging requires the application of a curvature algorithm along the energy direction to extract clear band dispersions comparable to room temperature data, which was treated using the same algorithmic settings[41]. The original data at both 300 K and 10 K are presented in Supplementary Figure S7. For the room temperature data, presented in Figure 4A, tracing of the band dispersion (dashed white lines) provides a reference template. Superimposed onto the band structure at 10 K, displayed in Figure 4B, the data reveal substantial similarity despite reduced spectral quality. The strong correspondence between 300 K and 10 K band structures, beyond minor energy shifts reported in similar layered materials[42,43], supports the validity of comparing calculations performed for antiferromagnetic order with ARPES data acquired in the paramagnetic phase. Further, this finding suggests that magnetic order induces only minimal changes to the overall electronic band structure in $CrPS_4$.

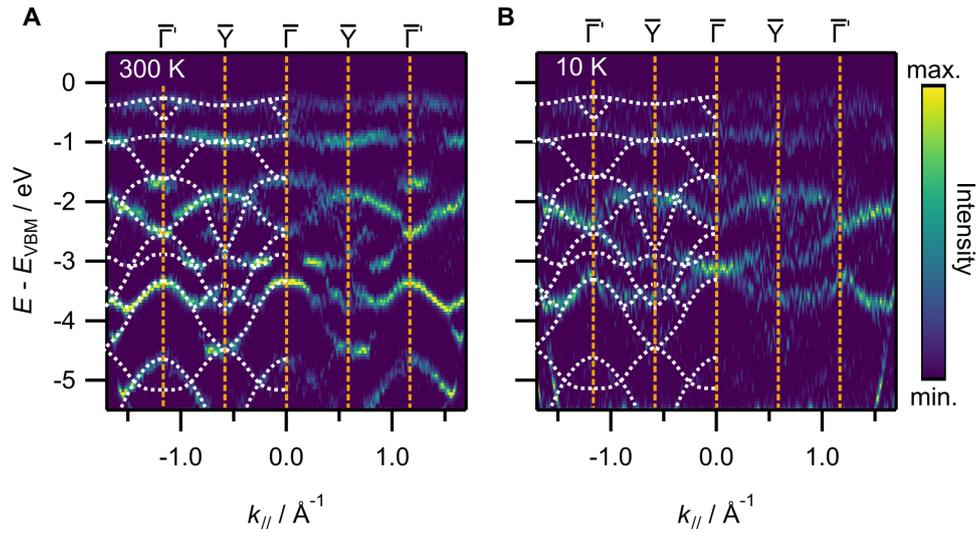

**Figure 4. Band Structure in the Para- and Antiferromagnetic Phase.** Curvature plots of the experimental band structure along the $\overline{Y}$-$\overline{\Gamma}$-$\overline{Y}$ direction at 300 K **A** and 10 K **B**, respectively. For the 300 K data, bands were traced by dashed white lines, which, when superimposed on the 10 K data, highlight the qualitative similarity of the two phases. The data was acquired using linearly-polarized light with a photon energy of $h\nu$ = 21.21 eV.

*Hybridization Effects in the Cr 3d-Manifold:*

The $Cr^{3+}$ ions in $CrPS_4$ are centered in slightly distorted S octahedra, being subjected to a CF that splits the five-fold degenerate 3d orbitals into a lower-energy $t_{2g}$ orbital triplet and a higher-energy $e_g$ doublet. For the $Cr^{3+}$ ion in its $3d^3$ configuration, Hund's rule filling results in one spin-up electron occupying each of the three degenerate $t_{2g}$ orbitals, with the two $e_g$ orbitals remaining nominally unoccupied. While this splitting defines a half-filled $t_{2g}$ configuration with parallel spins, the actual ground state is further shaped by Hund's exchange $J_{ex}$, which aligns spins within the $t_{2g}$ manifold, and the on-site Coulomb repulsion $U$. Crucially, the extent to which these orbitals then hybridize with the surrounding S ligands determines whether the electronic ground state remains predominantly localized and magnetic, or becomes substantially modified by covalent bonding. Understanding this hierarchy — which orbitals couple, how strongly, and with what consequence — is key to connecting magnetism, optical activity, and electronic transport in this layered semiconductor.

Figure 5 disentangles this problem at the orbital level. For simplicity, here we restrict the discussion to three representative cases: the Cr $d_{xy}$ orbital as an exemplary $t_{2g}$ state, the Cr $d_{z^2}$ orbital as an $e_g$ state, and the S $p_z$ orbital as a key ligand state (a full comprehensive summary including all various orbital contributions is presented in Supplementary Section S8). Figure 5 presents both the pDOS and momentum-resolved spectral weights, along with Wannier functions to visualize orbital character and spatial overlap.

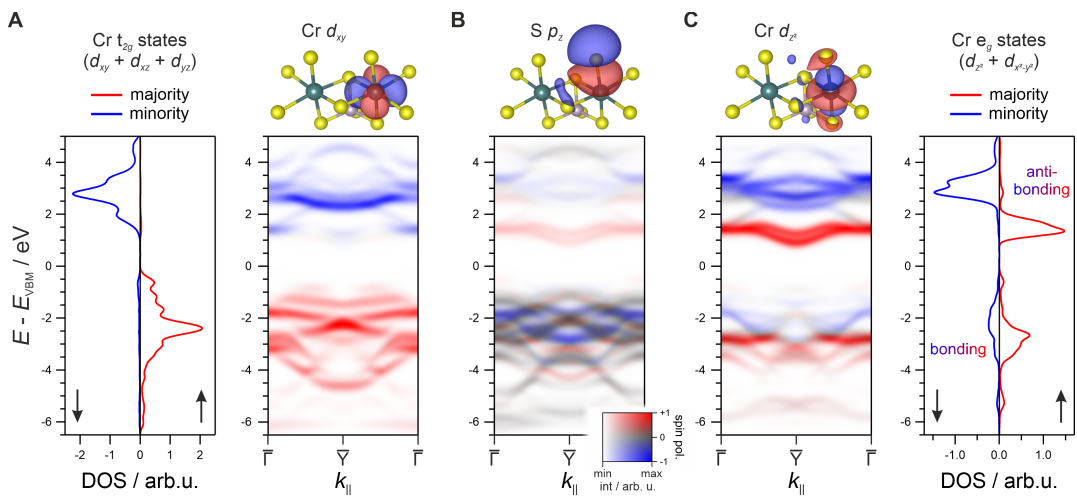

**Figure 5. Orbital-Selective Hybridization in $CrPS_4$. A** Left: pDOS of the Cr $t_{2g}$ manifold showing clear spin splitting (red = majority, blue = minority). Right: momentum-resolved spectral weight of the Cr $d_{xy}$ state (representative $t_{2g}$ orbital) with the respective Wannier function, displaying weak dispersion and localization on the Cr site. **B** Spin-resolved pDOS of S $p_z$ shows intermixed spin character in the occupied region but a rigid 1–2 eV spin splitting in the unoccupied region. **C** Left: momentum-resolved spectral weight of the Cr $d_{z^2}$ state (representative $e_g$ orbital) with the respective Wannier function showing significant weight on S ligands. Right: the $e_g$ orbital's pDOS reveals antibonding (unoccupied, blue) and bonding character states (valence, red). The 2D color scale indicates spin polarization (red/blue) and intensity (grey), and applies consistently across all momentum-resolved panels.

From the pDOS of the $t_{2g}$ states, displayed in Figure 5A, a clear spin splitting is evident: the majority states are fully occupied, while the minority states remain fully unoccupied. The momentum-resolved spectral weight, exemplified by the $d_{xy}$ state, reveals a variety of bands with weak to intermediate dispersions. Crucially, the lobes of the $d_{xy}$ orbital point in between the S atom positions, resulting in minimal direct overlap with the S 3p states. This geometric constraint yields a strong atomic orbital-like character, as illustrated by the corresponding Wannier function — the orbital weight remains tightly localized on the Cr site. This weak covalency, realized in a $\pi$-like bonding geometry, preserves the atomic-like spin polarization expected for a $3d^3$ ion. In turn, the $t_{2g}$ orbitals primarily stabilize the local magnetic moments while contributing only marginally to chemical bonding. The spin-resolved bands of the Cr $t_{2g}$ states show a clear separation into majority and minority channels. In sharp contrast, the occupied S $p_z$ bands (Figure 5B) exhibit intermixed majority and minority states, appearing mostly grey in the 2D color scheme with occasional blue and red streaks. This mixing in the valence region already hints at orbital coupling despite the weak $t_{2g}$-$p$ overlap. In the unoccupied region, however, a distinctly different picture emerges: a rigid offset between majority and minority states of approximately 1–2 eV is now clearly evident. Notably, comparison with the Cr $d_{z^2}$ projected DOS (Figure 5C) reveals that this splitting is a direct consequence of strong hybridization between the S $p_z$ and Cr $d_{z^2}$ orbitals. This coupling becomes particularly evident in the unoccupied region, where $e_g$ states are expected to have their maximum contribution. A pronounced spectral overlap between the S $p_z$ bands and the Cr $d_{z^2}$ bands in this energy window clearly reflects the strength of the hybridization resulting from $\sigma$-bond formation. In line with this, the Wannier function associated with the Cr $d_{z^2}$ orbital exhibits substantial weight on the S ligands along the direction of the $d_{z^2}$ orbital lobes — a direct spatial signature of hybridization.

Interestingly, the $d_{z^2}$ state also shows non-negligible spin-polarized weight in the occupied region, deviating from the simplified view of empty $e_g$ states. Within the $e_g$-pDOS, two distinct regions of intensity emerge. The dominant contribution lies in the unoccupied region, which can be attributed to antibonding states formed between the Cr $e_g$ and S $p$ orbitals, while the damped peaks below $E_{VBM}$ can be interpreted as their bonding counterparts. This orbital-selective hybridization is also evident in the S $p_z$ channel, which mirrors the dispersion of the $d_{z^2}$ bonding states, confirming their shared orbital character. The resulting admixture of orbitals within the $e_g$–$p_z$ manifold has a key consequence: it relaxes the dipole selection rules that would forbid pure $d$–$d$ transitions[44,45], partially rationalizing the strong sub-gap optical absorption characteristic of CrPS$_4$[22,23].

Overall, a clear dichotomy emerges from this orbital-level analysis: the localized, spin-polarized $t_{2g}$ orbitals sustain magnetism with minimal bonding character, whereas the strongly hybridized $e_g$–S $p$ combinations mediate both covalent bonding and optical coupling. The coexistence of these two regimes — localized exchange and extended covalency within a single 3$d$ shell — is decisive for the peculiar correlated electronic landscape of CrPS$_4$.

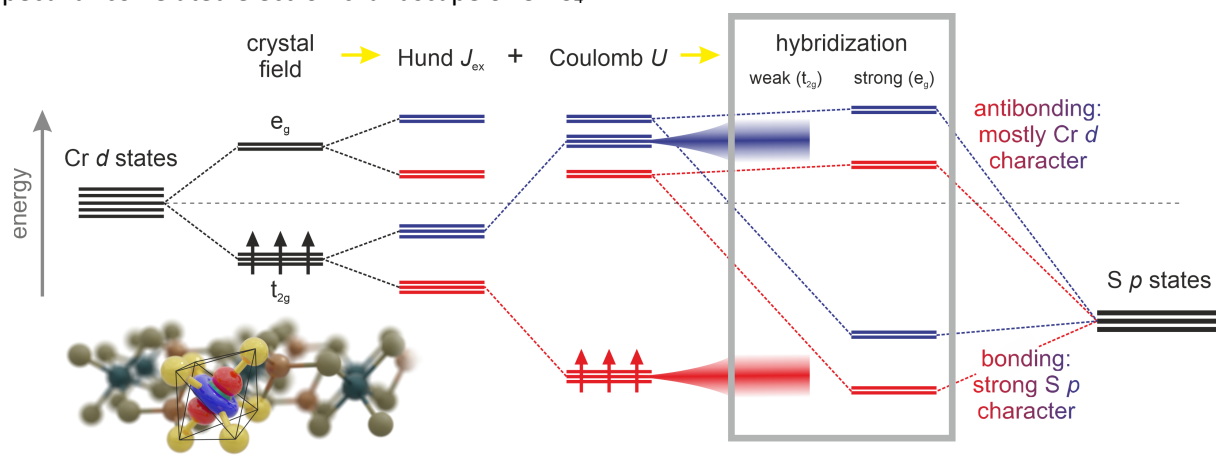

**Figure 6. Conceptual Hierarchy of Interactions in CrPS$_4$.** Schematic evolution of the electronic structure from atomic $d$ orbitals (left) through successive symmetry-breaking and correlation-driven steps. The octahedral CF splits the $d$-manifold into $t_{2g}$ and $e_g$ levels. Hund's exchange and Coulomb repulsion align spins and penalize double occupancy within the $t_{2g}$ states, yielding a spin-polarized half-filled configuration (indicated by arrows). Hybridization with ligand (S) $p$ orbitals then produces orbital-selective coupling: weak $t_{2g}$–$p$ hybridization causes moderate broadening, while strong $e_g$–$p$ interaction generates separated bonding (lower energy, strong S $p$ character) and antibonding (higher energy, mostly Cr $d$ character) pairs. The resulting $e_g$–$p$ mixing enables optical absorption through spin admixture, while $t_{2g}$ magnetism remains robust. Inset: CrPS$_4$ with a highlighted Cr atom (green) and its $d_{z^2}$ orbital surrounded by a distorted S (yellow) octahedron; phosphorus atoms are shown in orange.

Figure 6 condenses this hierarchy into a conceptual scheme. Starting from the fivefold-degenerate atomic $d$ levels, the electronic structure evolves through a sequence of symmetry-breaking and interaction-driven steps: the octahedral CF splits the $d$ states into $t_{2g}$ and $e_g$ subsets; Hund's exchange and Coulomb repulsion stabilize spin alignment and penalize double occupancy within $t_{2g}$ orbitals; ligand hybridization then mixes the $e_g$ and S $p$ orbitals into separated bonding and antibonding pairs, while weaker hybridization between $t_{2g}$ and S $p$ states results in moderate broadening. The displayed progression: **CF → Hund $J_{ex}$ + Coulomb $U$ → hybridization** provides a convenient conceptual framework for understanding the Cr–S interaction, even though in the real material these contributions act concurrently. In the antiferromagnetic ground state, Cr hosts robust local moments rooted in weakly hybridized $t_{2g}$ states; although the bulk valence band is not net spin-polarized, local spin polarization is strong. By contrast, the optical response is governed by in-gap excitations that become weakly dipole-allowed through $e_g$–ligand $p$ hybridization.

**Conclusion**

In conclusion, this study provides the first experimental determination of the electronic band structure of CrPS$_4$ by means of momentum-resolved photoemission spectroscopy, both below and above its Néel temperature. This establishes a comprehensive experimental benchmark that further validates theoretical simulations, revealing distinct hybridization behaviors of the Cr 3$d$ $t_{2g}$ and $e_g$ orbitals with S $p$ states.

Our ARPES measurements in both paramagnetic (300 K) and antiferromagnetic (10 K) phases, complemented by DFT+U calculations, reveal excellent agreement between experiment and theory, establishing DFT+U as a reliable computational framework for investigating CrPS$_4$. The valence band region is dominated by a rich mixture of Cr 3$d$ and S 3$p$ bands, with the band gap exhibiting ligand-to-metal charge-transfer character. Comparison of the band structure between the para- and antiferromagnetic phase reveals no major differences in band structure symmetry or band dispersion, as expected for layered A-type antiferromagnets. To identify more nuanced changes in the band structure induced by magnetic ordering, further progress in the preparation of the strongly charging semiconductor and experiments capable of optically identifying few-monolayer samples in UHV will be necessary.

DFT+U calculations focusing on the octahedrally coordinated Cr 3$d$-manifold reveal fundamentally different behaviors for the $t_{2g}$ and $e_g$ orbitals. The $t_{2g}$ orbitals, undergoing only weak $p$-$d$ hybridization, follow a CF plus Hund's-rule picture, producing fully spin-polarized bands that drive magnetic properties within one layer. Conversely, $e_g$ orbitals undergo strong hybridization with S $p$ orbitals, deviating from the simplified CF/Hund scenario and acquiring finite spectral weight in the occupied valence band region. The strong ligand–metal mixing can relax otherwise strict dipole selection rules, enabling and enhancing nominally forbidden $d$–$d$ transitions that shape the optical response of CrPS$_4$. Thus, the differing hybridization characters of $t_{2g}$ and $e_g$ states allow the Cr 3$d$ manifold to play two complementary roles — stabilizing magnetic order while governing key aspects of optical behavior, both crucial for prospective applications.

Building on these insights, future studies should target the Cr 3$d$ states directly—using spin-resolved ARPES to map their spin texture and controlled doping (chemical or electrostatic) to tune 3$d$ occupation and $p$–$d$ hybridization. In parallel, probing the dominant $d$–$d$ transitions with orbital-selective spectroscopies such as resonant inelastic X-ray scattering, complemented by momentum- and time-resolved photoemission, will quantify how ligand–metal mixing governs these excitations across energy, momentum, and time.

**Methods and Experimental Details**

*Sample Preparation:*

Thin flake samples were prepared on a gold thin-film substrate following the method presented by Watson et al.[32]. First, a 100 nm gold layer was thermally evaporated onto a natively oxidized Si chip under high-vacuum conditions (base pressure $\approx 5 \cdot 10^{-6}$ mbar). The chip was then glued to a copper (Cu) sample plate, with the gold side facing down, using conductive silver epoxy (EPO-TEK® H21D Electrical Cond. Adhesive). Inside a glovebox with inert $N_2$ atmosphere ($O_2$ < 0.2 ppm, $H_2O$ < 0.03 ppm), the Si chip is detached from the Cu plate, leaving behind a smooth and atomically clean gold surface. Subsequently, a commercial $CrPS_4$ bulk crystal (hqGraphene) was cleaved using Kapton tape in the inert atmosphere, and the tape with the cleaved crystal was immediately pressed onto the pristine gold surface. Finally, the system is transferred under $N_2$-atmosphere into an ultra-high-vacuum (UHV) chamber, where the tape is removed at a pressure of $4 \cdot 10^{-8}$ mbar leaving behind the flakes, investigated using μ-ARPES.

*Momentum-Resolved Photoemission Study:*

This experimental setup[46] consists of a KREIOS 150 MM momentum microscope (Specs GmbH), which operates at a base pressure of $3 \cdot 10^{-10}$ mbar and is capable of performing PEEM as well as ARPES in the framework of momentum microscopy (MM). In PEEM mode, the microscope images the real-space distribution of photoemission intensity, over a 150 μm diameter circle, in dependence of the photoelectron's kinetic energy *E* filtered by a hemispherical analyzer. In combination with the difference in work function between materials, this allows for real-space imaging of samples on the microscale. In momentum mode, the microscope captures the photoemission intensity in 2D momentum space at a given *E*, so-called momentum maps, with an approximate field of view of $k_x$, $k_y \in [-2.0, 2.0]$ Å$^{-1}$. Scanning *E* yields a stack of momentum maps, which can be cut along arbitrary directions to obtain the electronic band structure. Integrating the photoemission intensity over the entire momentum FOV yields the EDC, which approximately corresponds to the DOS. Apertures in the PEEM column allow us to select specific regions of the sample in real space, enabling μ-ARPES measurements on individual flakes. The microscope is coupled to a monochromatized helium (He) gas discharge lamp, which is tuned to the He I$\alpha$ line, providing *p*-polarized 21.21 eV photons incident at 68° relative to the sample surface normal. Cooling to cryogenic temperatures is performed using a liquid He flow-cryostat.

*DFT+U Calculations:*

The experimental results are supplemented by DFT+U calculations carried out with the Quantum ESPRESSO suite (version 6.7MaX)[47–49]. Exchange-correlation effects were treated within the generalized gradient approximation using the Perdew-Burke-Ernzerhof (PBE) functional[50], together with scalar-relativistic PAW (projector-augmented-wave) datasets[51–53]. Grimme's D3 dispersion corrections were included to account for vdW interactions between stacked layers[54]. The kinetic-energy cutoffs were 80 Ry (wavefunctions) and 800 Ry (charge density). Structural relaxations used the Broyden-Fletcher-Goldfarb-Shanno (BFGS) algorithm until atomic forces were smaller than 1×10$^{-3}$ Ry/au, while the convergence threshold for the total energy was 1×10$^{-4}$ Ry. The BZ was sampled with Γ-centered *k*-meshes of 5×5×3 during relaxation, 9×9×5 for electronic calculations, and 13×13×5 for non-self-consistent runs. Additional calculations were carried out for a rotated unit cell (9×9×3 *k*-mesh) in which the octahedral axes of the Cr-centered octahedron are aligned with the Cartesian axes to disentangle orbital contributions. The rotated cell was re-relaxed on a 5×5×3 grid to mitigate minor numerical artifacts introduced by the rotation. All calculations used Gaussian smearing with σ = 0.05 eV for self-consistent and non-self-consistent steps, consistent with prior MPX$_3$ studies[55,56]. For direct comparison to photoemission intensities, the DOS was further broadened by 0.2 eV. The Cr 3$d$ states were treated within the DFT+U framework using the Dudarev scheme with an effective $U_{\text{eff}}$ of 2 eV ($U_{\text{eff}} = U$-$J$), adopted from Susilo et al.[23], which reports good agreement with experimental results. The DFT+U band structure was projected onto a basis of maximally localized Wannier functions using the Wannier90 code[57–59]. The Wannierization included Cr $d$ and P, S $p$ states to accurately describe the occupied valence band region. The accuracy of the Wannier interpolation was verified by comparing the Wannier-interpolated band structure with the original DFT+U results in Supplementary Section S9. Assuming minimal out-of-plane dispersion[32], we utilized the WannierTools package[60] to simulate theoretical 2D momentum maps at $k_z = 0$, for comparison with experimentally obtained momentum microscopy data. The maps were computed with a broadening parameter of $\eta_{\text{arc}} = 0.2$ eV (corresponding to a Lorentzian FWHM of 0.4 eV). This value was chosen to reproduce the experimentally obtained maps, effectively approximating the combined lifetime, as well as instrumental and thermal broadening in a single parameter. Wannier functions of the rotated system were further inspected to confirm the orbital character, in particular of the Cr $d$ states. The unit cell contained 24 atoms corresponding to two stacked layers. To capture the experimentally relevant antiferromagnetic ordering, the initial spin configuration was set with opposite orientations in the adjacent layers.

*AFM, PL- and Raman-Spectroscopy:*

For structural characterization of the measured flake, an alpha300 RA system (Witec GmbH), which provides high-resolution optical imaging, AFM, and Raman spectroscopy, was employed. All measurements were performed under ambient conditions and at room temperature, with the AFM scans in contact mode and the Raman spectra using a 488 nm CW laser at 1 mW. Here, the optimal power, still avoiding radiation damage, common to CrPS$_4$[21], was determined by observing damage effects at different powers, as presented in the Supplementary Section S2.

**Resource Availability**

*Lead contact*

Further information and requests for resources and reagents should be directed to and will be fulfilled by the lead contact, Lasse Sternemann (lasse.sternemann@tu-dortmund.de).

*Materials availability*

This study did not generate new unique materials.

*Data and code availability*

All optical images supporting the findings of this paper are included within the article or any supplementary files. The raw ARPES, AFM, and Raman spectroscopy data are available in the ZENODO database. Any additional information required to reanalyze the data reported in this paper is available from the lead contact upon request.

**Acknowledgments**

The momentum microscope was financed by the DFG through the project INST 212/409 and by the "Ministerium für Kultur und Wissenschaft des Landes Nordrhein-Westfalen". We also acknowledge financial support by the DFG through project dd2D (project no. 55818086; CI 157/11-1) as well as support from the European Union's Horizon 2020 Research and Innovation Program under Project SINFONIA, grant 964396. The AFM/Raman setup has been financed by the DFG (project no. 507298853) and MKW.NRW. Leon Becker and Stefan Tappertzhofen acknowledge funding from the DFG (project no. 524569125).

The authors express their gratitude to Dirk Schemionek for his assistance during the preparation process.

**Author contributions**

**L.S.**: Conceptualization, Investigation – Sample Preparation, ARPES, AFM, Raman, Writing – original draft, Writing – review & editing, Visualization. **D.M.J.**: Conceptualization, Investigation – DFT+U, AFM, Raman, Writing – original draft, Writing – review & editing, Visualization. **E.B.**: Investigation – ARPES, Writing – review & editing. **R.L.**: Investigation – Raman Spectroscopy, Writing – review & editing. **J.E.N.**: Investigation – ARPES, Writing – review & editing. **M.M.**: Writing – review & editing. **L.B.**: Investigation – AFM, Raman, Writing – review & editing. **A.C.A.**: Investigation – Raman Spectroscopy, Writing – review & editing. **F.A.**: Writing – review & editing. **S.T.**: Funding acquisition, Writing – review & editing. **M.C.**: Conceptualization, Funding acquisition, Writing – review & editing, Supervision.

**Declaration of interest**

The authors declare no competing interests.

**Declaration of generative AI and AI-assisted technologies in the writing process**

During the preparation of this work, the authors used ChatGPT to refine the manuscript regarding grammar, readability, and clarity. After using this tool/service, the authors reviewed and edited the content as needed and take full responsibility for the content of the publication. No additional use of artificial intelligence was made during data acquisition, analysis, interpretation, or other aspects of this study.